# Repurposing of Ring Framework through TR screening

Surendra Kumar[‡], Cheongyun Jang[‡], Lalita Subedi, Sun Yeou Kim, Mi-hyun Kim[*]

Gachon Institute of Pharmaceutical Science & Department of Pharmacy, College of Pharmacy, Gachon University, 191 Hambakmoeiro, Yeonsu-gu, Incheon, Republic of Korea

[‡]The authors are co-first authors
[*]Author for correspondence
E-mail: kmh0515@gachon.ac.kr

**Abstract**

In the drug repurposing approach, the chemically diverse and potentially safe molecules can be explored as therapeutic potential out of those originally targeted indications. However, the intellectual property rights, and competitive over-heating issue from sharing the same drug pipeline information are drawback of the approach. In current paradigm of drug discovery, despite approval of new chemical entity, the finding of a new ring system for drug discovery have stagnated. In addition, there are still rarely used ring systems from FDA approved drug that can be optimized and used for repurposing purpose. In the present work, these rarely used drug ring system were re-purposed with our bidirectional target and ring screening (TR-dual screening) approach within our defined objectives. The TR-dual screening suggested RARbeta and cyproheptadine as a pair of target and ring system. The selected ring system was virtually grown with predefined criterion, ranking of molecular docking and top score poses made us select test compounds for synthesis and biological proof. Among the tested compounds, **6c** proved RARbeta agonist activity. The concept of TR dual optimization can contribute to picking either a compound having acceptable ADMET profile or a target for a selected drug scaffold.

**Introduction**

The paradigm of current drug R&D is 'selection & concentration' and current massive investment in drug R&D has not been just 'massive' but 'efficiently massive' such as described in 5R-framework of Astrazeneca(1). In particular, 'Go/No Go' decision in R&D stages means the willing to choose a more promising pipeline within available resource(2). However, the strategy can produce neither satisfied success rate nor a chance investigating black box information of the pipeline during R&D. Data generated at drug R&D process can't perfectly support assumed proof of concept (POC) to show the limitation in controlling hidden black box eg., exposure concentration of a drug near to a therapeutic target, POC, therapeutic window in clinical level (1). To overcome the productivity crisis, drug repurposing approaches have been developed. The chemically diverse and potentially safe molecules can be explored as therapeutic potential out of those originally targeted indications. However, the intellectual property rights, and competitive over-heating issue from sharing the same drug pipeline information are drawback of the approach. The described status of drug R&D made us highly motivated in thinking how to control the black box for overcoming current failure of drug R&D.

 Even though any researcher surely can't propose a breakthrough paradigm or strategy for controlling the black box, trials discriminative from current approaches are required and someday researchers can find any breakthrough or advance from the accumulated trials. As one discriminative drug discovery approach, our research group proposed 'chemistry-oriented synthesis (ChOS)', the approach starting drug discovery "from an unprecedented drug scaffold" rather than "from a specific target"(3, 4). Through the ChOS approach, drug discovery driven by the novelty of a drug scaffold can compensate the general approach driven by a specific target to broaden the drug space of artificial drugs(4). In the view of 'selection & concentration', general approach focuses on the 'target' and ChOS approach focuses on 'drug'. After the study, now we really wonder transferring the point of view can make our discriminative insight and methods close to black box information of drug R&D. In the history of drug discovery, until 1980, synthetic chemists acquired novel drug scaffolds through serendipitous findings & intensive investigation during their active synthesis(5, 6). However, since the golden age, in the spite of annual approval of around 10 NMEs, it was reported that developing an unprecedented drug scaffold or a novel drug ring system, and finding a new ring for drugs have stagnated(7). In the analysis on frameworks/ring systems/rings of FDA approved drugs, 237 ring systems were very rarely used (less than three times) among every ring system in FDA orange book. From the report, we thought (1) how much a ring system can make an effect on drug-like property of a whole drug structure, (2) how about repurposing a successful ring system from the indication of its approved drug to another indication, and (3) how to elicit a new promising drug scaffold from the successful ring system with desirable PK property. Even though a research corresponding to the described questions can't be an efficient strategy in drug R&D now, data generated from the research can be discriminating with the data driven from general approach. In this study, we chose one ring system through our bidirectional screening, 'target (T) & ring system (R) dual screening: TR- dual screening', generated virtual library of the chosen ring system, conducted chemo-centric target screening of the virtual library, and repurposed the chosen ring system from histamine receptor (anti-allergic) to retinoic acid receptor *beta* (anti-cancer/inflammation) with DMPK measurement.

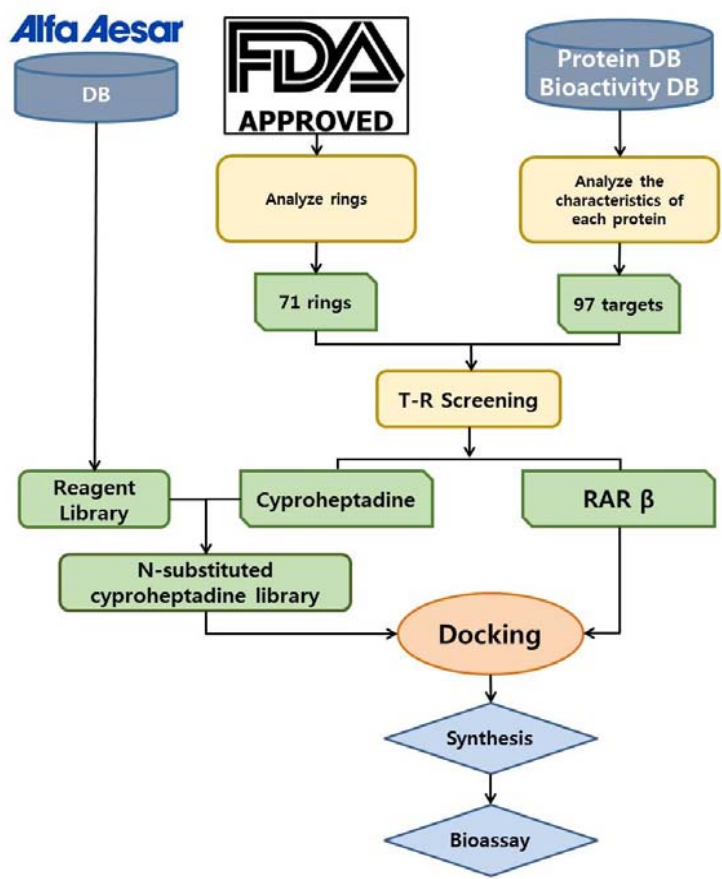

**Figure 1:** The general workflow for Target (T) – Ring (R) screening.

**Results**

**Descriptor selection for ring system repurposing:**

For the scalable & bi-directional TR screening, descriptors were assigned to reported ring systems (349) and the descriptors were used to reduce the number of ring system. Properties of 349 ring systems extracted from every FDA-approved drug in Orange book (7) were calculated through CDK toolkit to analyze (1) fragment complexity, (2) hydrogen bond acceptors, (3) hydrogen bond donors, (4) molecular weight, and (5) VABC volume descriptor [ref]. Among all the properties calculated, VABC descriptor represents the approximate van der Waals volume ($Å^3$/molecule) of molecules and can be calculated by the following formula:

$V_{vdW}= \sum$ all atom contributions $- 5.92N_B - 14.7R_A - 3.8R_{NR}$

$N_B$ is the number of bonds, $R_A$ is the number of aromatic rings, and $R_{NA}$ is the number of nonaromatic rings. From 349 ring systems with less than three frequencies in clinically approved drug, 114 ring systems were chosen based on VABC Volume Descriptor > 140 (8) and 71 ring systems remained after the summation of HA+HB < 3. In this study, 3D structure of a core (a ring system) rather than side chains (substituents) was focused for maximizing the effect of a ring system on drug-target interaction. According to the criteria [Frequency in Orange book ≤ 2, MW ≤ 500, HA+HD ≤ 2, VABC ≥ 140] in the **Figure 2**, 71 ring systems were selected from total 349 ring systems (**see supplementary file1; Table S1**).

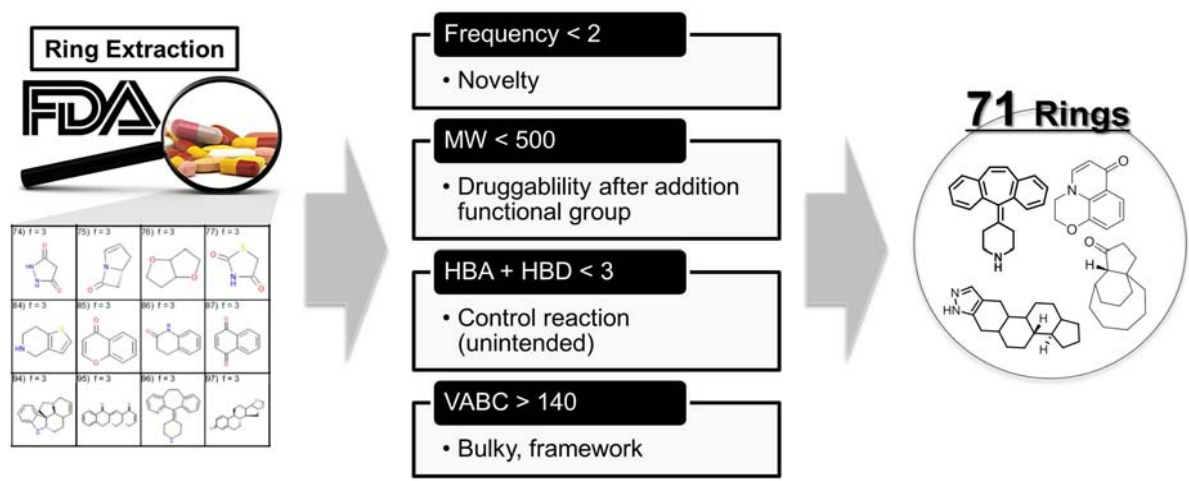

**Figure 2:** Ring reduction workflow.

**Target selection for ring system repurposing:**

In PDB bank, 38,529 targets were filtered under the conditions of (1) more than five available PDBs of a target, (2) existence of a ligand in the PDBs, (3) carbon construction of a ligand in the PDBs, and (4) desirable range of molecular weight (250 to 800) to produce 1,714 targets. The 292 targets were common in TTD database (3,261 targets), and ChEMBL version 21 (6,930 targets), and the refined PDB (1714 targets) so that every ChEMBL ligands of the 292 targets were collected and manipulated. Among 292 targets after the manipulation, the number of target was reduced into 97 targets (**Figure 3**).

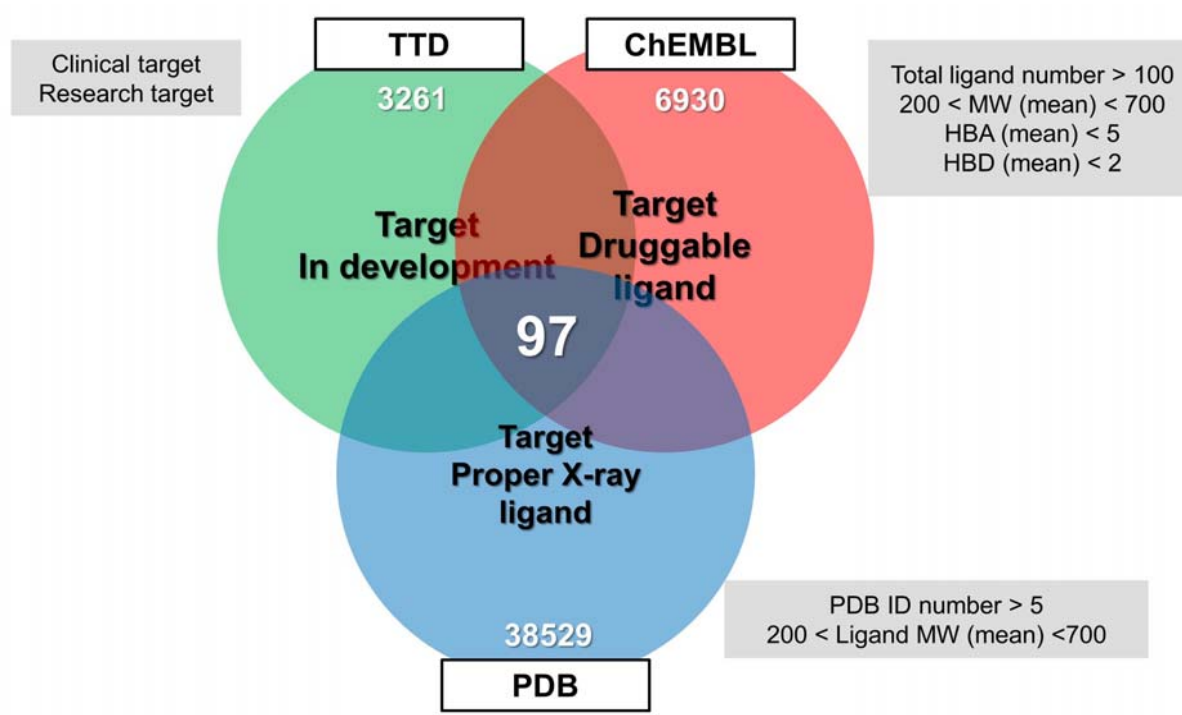

**Figure 3:** Target reduction workflow.

**Target (T)-Ring (R) screening Result:**

The 3D similarity calculation between 71 ring systems (query) and 3424 ligands of PDBs produced 246,512 paired data with several scores from 3 metrics and 2 color/shape (**Figure 4**). Expectedly, the complexity of the bi-direction screening as well as narrow distribution and deviation of the 3D similarity scores couldn't make us choose one 'target-ring system pair'. In sequence, we chose the best target against each ring system (97 rows, **see supplementary file1; Table S2**) and the best ring system against each target (71 rows, **see supplementary file1; Table S3**) instead of choosing one uncertain best pair. PDBs annotated in the two best lists (best target, best ring system) were collected and after the annotation of accumulated count of duplicated PDBs, the duplicated was removed to present 131 PDBs (97 target) for second screening (Docking based TR screening).

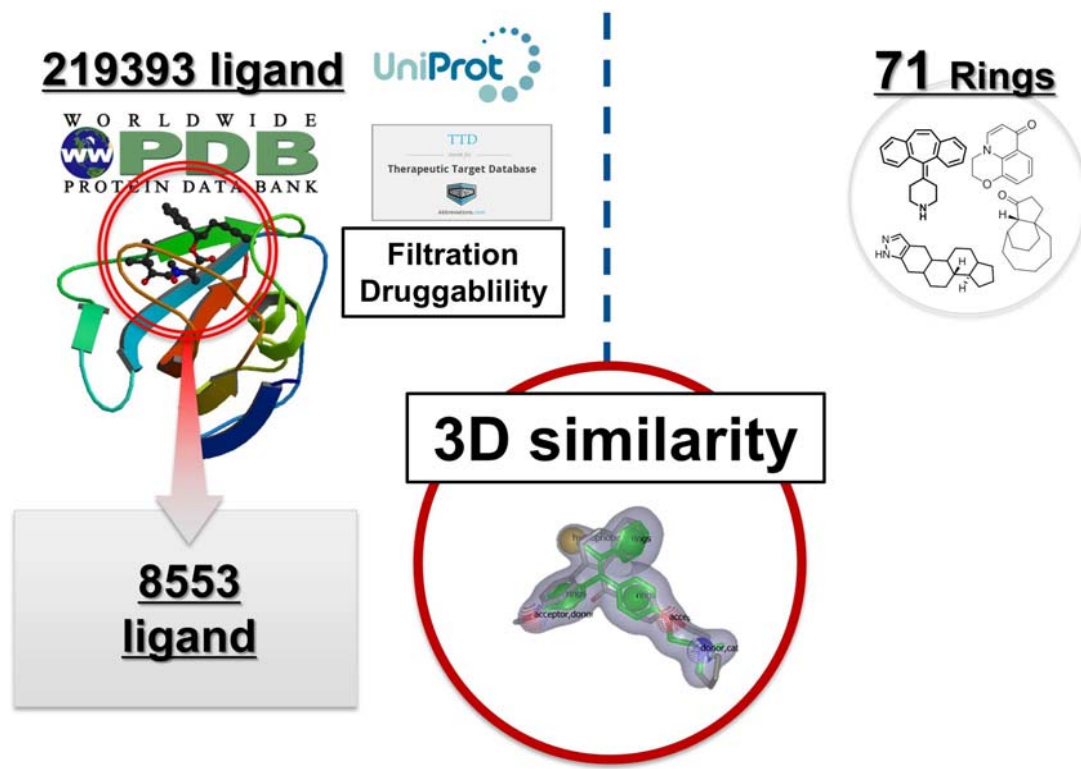

**Figure 4:** 3D-similarity calculation between x-ray ligands and Rings.

**Docking based TR screening result:**

To select the best pair of target and rings, a second screening comprised of molecular docking was performed on 71 ring (R) system and 131 PDBs (97 target (T) ). Due to missing atom types for ring (347-349_2) system, it was excluded during docking. The Standard Precision (SP) and Xtra Precision (XP) mode is employed during the binding mode search. After docking, Uniprot accession id was assigned to PDBs and single entry among homologous protein was retained. A matrix of (97 row x 70 column) was prepared **(see supplementary file 2)** that consists of docking score values. The docking score values for each targets and each rings were transformed into ranking values and heatmap based on ranked targets and rings were created. The heatmap present the data in different colour intensity with high ranked as dark blue and low with yellow color. The heatmap for ranked target and ranked ring is shown in **Figure 5 and 6**.

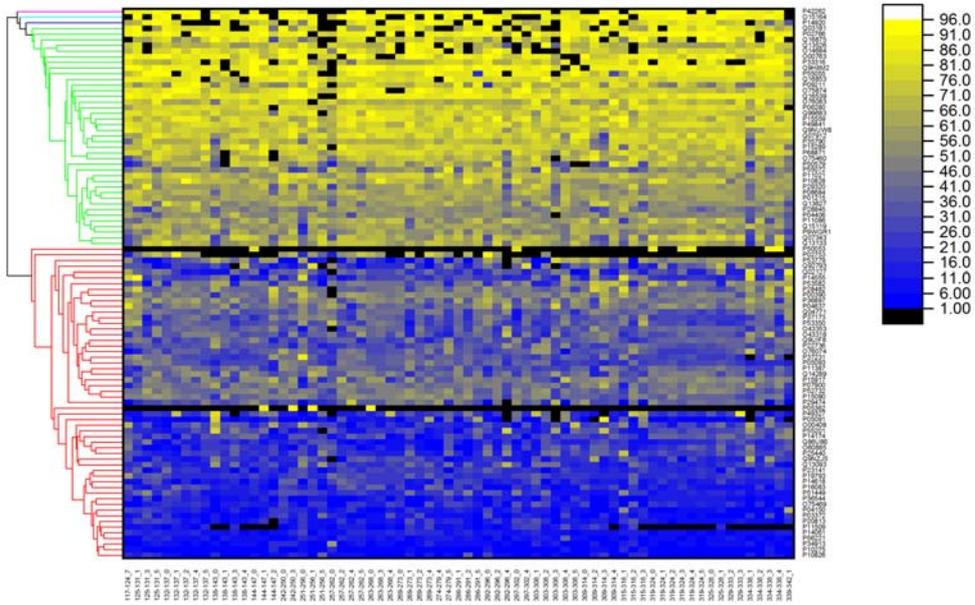

**Figure 5:** Heatmap based on target (97) ranking according to docking score against Rings.

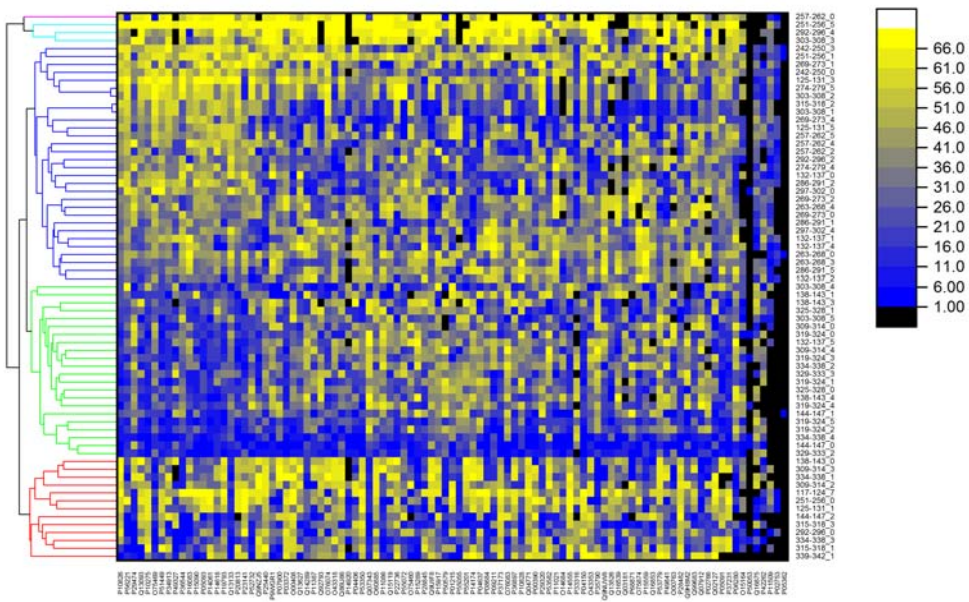

**Figure 6:** Heatmap based on Ring (69) ranking according to docking score against target.

**Overview on the selected target and ring system:**

The selected target P10826 belongs to the family of nuclear receptor denoted as RAR β (Retinoid Acid Receptor Beta). The natural modular of this receptor is retinoic acid, which specifically binds to retinoid acid receptor and then forms heterodimers with other receptors such as estrogen receptor alpha and AP-1 receptor and further activates downstream to regulate cell differentiation(9, 10). The retinoic acid receptor is divided into three classes of alpha, beta, and gamma, of which beta has been shown to have anti-cancer effects in epithelial cells. Multiple studies reported that RARβ play a role in breast cancer cell metastatic process and found to interact with ATRA (all trans retinoic acid)(11). In addition, the neuroprotective effect of retinoic acid as well as cancer cells is now being revealed(12). The selected ring system (339-342_1) has scaffold that abundantly found in ant-histamine drugs and known as cyproheptadine(13) but never was considered for RAR β in literature study. The selected ring system is rigid without rotatable bond. The nitrogen as the hetero atom is unique and it is easy to introduce the functional groups.

**Virtual library generation and molecular docking:**

Since the x-ray crystal structure in 4JYG has carboxyl functional group and it makes H-bond with Ser280 and Arg269 amino acid residues of retinoic acid receptor, thus the reagent library with carboxyl functional was selected to create the virtual libraries. During the virtual library generation, the cyproheptadine was set to core structure in demethylation state and libraries were grown to amine position. Additionally, the druggability of every compound in the virtual library was predicted to filter out undesirable compounds according to the pre-defined criteria and 4954 compounds were retained.

These compounds were further docked into the selected target protein (PDB ID: 4JYG). During docking, the energy window for ring sampling was increased to enhance the conformer sampling and write out at most 2000 poses per ligand. From all the docking results, the poses were ranked according to docking scores and top scoring poses selected for further analysis. Comparing the binding mode of top ranked poses from virtual library and crystal bound ligand, a structural complementary in their binding mode was found (**Figure 7, Table1**). Among all the poses, compound 6c exhibit maximum binding affinity in terms of docking score of -15.568kcal/mol. Similarly, the compound 6b shows at par binding affinity of -15.324. In addition, compound 8a, 6d, 6e, 6a, shows lower binding affinity and 6f shows lowest binding affinity among selected poses.

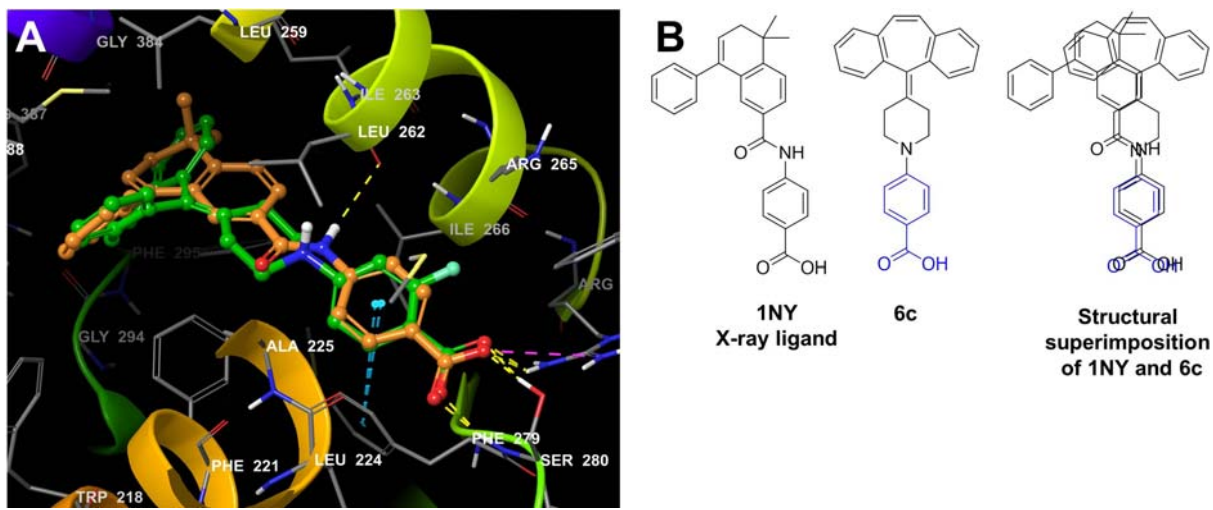

**Figure 7:** (A) The binding mode and superimposition of docked x-ray ligand and 6c (B) The structural superimposition of x-ray ligand and 6c

**Table 1: The selected N-arylated cyproheptadine derivatives.**

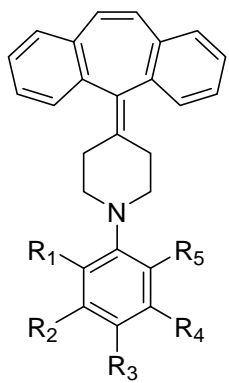

| Compound ID | $R_1$ | $R_2$ | $R_3$ | $R_4$ | $R_5$ |
|---|---|---|---|---|---|
| 6a | H | COOH | $CH_3$ | H | H |
| 6b | H | $CH_3$ | COOH | H | H |
| 6c | H | H | COOH | H | H |
| 6d | $NO_2$ | H | COOH | H | H |
| 6e | H | COOH | H | H | H |
| 6f | COOH | H | H | H | H |
| 8a | $NH_2$ | H | COOH | H | H |
| 8b | H | COOH | H | $NH_2$ | H |

**Discussion:**

The ring or ring systems plays an important role in medicinal chemistry either in the area of combinatorial chemistry, where ring systems are used as central scaffold or in scaffold hoping, where central core is replaced with bioisosteric ring system. In bioactive molecules, heteroaromatic ring contributes approximately 97% and least by aliphatic ring (65%). Almost 73.3% of bioactive molecules have fused aromatic ring system ranging from one to three simple five or six membered rings(14). Despite having such diversity, the substituents grown on these cores have drug like properties. Any drug repurposing methodology can be challenged by intellectual property rights of drug molecules that limit the re-purposed drug molecule for its reuse in different target indication. The challenges can be overcome by the proposed bi-directional target and ring screening methodology.

**Rings and Target Selection for repurposing:**

A total of 349 ring systems were extracted from every FDA approved drug in orange book and based on predefined filter 71 rings were obtained. These rings were diverse in nature ranges from two to six fused ring. These fused ring system, open wide possibility for site modification, however considering the synthetic accessibility can further guide for optimization with the domain of application. Similarly, from 38,529 targets using predefined filtering criteria a total of 292 targets (3424 proteins) were retained and these targets cover wide range of biological activity.

**Bi-directional Target (T)-Ring(R) screening:**

Due to large number of protein structures, the initial bi-directional Target-Ring docking was not feasible, so a 3D similarity score based screening was performed. The initial screening result produces two list (best target and best ring system), which were accumulated and after removing the duplicates 97 targets (131 PDBs) were retained. The reduced targets and rings were further subjected for second target and ring screening based on docking.

The bi-directional target and ring docking produces a matrix of docking score. Since our goal was to select best ring and target against each target and ring respectively, a heatmap was created. The heatmap has efficiently discriminated the poorly ranked targets for each rings and poorly ranked rings for each target, which were excluded from best pair selection criterion. After detailed analysis on two heatmaps (ranked targets and ranked rings), a matrix of best targets and best rings was created for selection of best pair of target and ring for repurposing. To select the best pair from matrix, the docking score is chosen as final selection criterion. After analysis of the docking matrix, it was found that, in most of the high docking score T-R pairs; the target has similar x-ray crystal ligand as with docked ring. Thus these pairs in spite of having higher docking score were excluded to reduce the biasness in selection of best T-R pairs. Based on final selection criterion, the best T-R pair of P10826 (target) and 339-342_1 (ring) was chosen for repurpose.

**Molecular docking of virtual library of cyproheptadine:**

The binding pocket of 4JYG is hydrophobic in nature, as majority of amino acid belongs to aliphatic or hydrophobic amino acid residues. Despite having hydrophobic interaction of the drug molecules, the binding affinity is further strengthen by H-bond from hydrophilic amino acid residues. The list of binding amino acid residues for selected poses is shown in **Table 2**. In the binding pose, the compound 6b, 6c both makes the π-π interactions with Phe279 and hydrogen bond with Arg269 and Ser280 amino acid residues respectively. Comparing the binding affinity with, compound 6d and 8a, it is revealed that, despite having same functional group on the core, these compounds have lower binding affinity. The nature of binding pocket with in 3Å is Aromatic hydrophobic (Phe192, Trp218, Phe221, Phe279, Phe295), Aliphatic hydrophobic (Leu224, Ala225, Leu259, Leu262, Ile263, Ile266, Leu298, Ile380, Val288, Leu391, Ile403, Met407), Polar, uncharged (Cys228, Ser280) and positively charged (Arg269) amino acid residues. The hydrophobically enriched binding pocket is complementary with bound ligand and imparts the binding affinity, however, presence of substituents and its orientation contributes to the differences in binding affinity observed. Likewise, the carboxylic acid at R3 position imparts better binding affinity as compound containing such groups makes H-bond with Arg269 and Ser280 that stabilizes in the binding pocket. Changes in the position of carboxylic acid, greatly reduces the binding affinity (i.e. compounds 6a, 6e, and 6f). Moreover, the presence of polar groups like $-NO_2$, or $-NH_2$ at R1 position did not increase the binding affinity. Similarly, the low binding affinity has been shown by compound 6f, which contains carboxylic acid at the same position. Thus, it is quite evident from docking studies that, presence of hydrophobic groups increases the binding affinity, that further strengthen by the suitable positioning of hydrophilic groups.

**Table 2: The binding affinity of synthesized N-arylated cyproheptadine derivatives against RAR β target protein (PDB ID: 4JYG).**

| Compound ID | Docking Score | Amino Acid Residues within 3 Å | Amino Acid Residues involved in interaction |
|---|---|---|---|
| 6a | -11.506 | Phe192, Trp218, Phe221, Leu224, Ala225, Cys228, Arg265, Ile266, Arg269, Phe279, Ser280, Gly294, Phe295, Leu298, Gly384, Val388, Leu391, Arg387, Met399, Ile403, Met407 | Phe279 (π-π) |
| 6b | -15.324 | Phe192, Trp218, Phe221, Leu224, Ala225, Cys228, Leu259, Leu262, Ile263, Arg265, Ile266, Arg269, Phe279, Ser280, Gly294, Phe295, Leu298, Ile380, Val388, Leu391, Met399, Ile403 | Arg269 and Ser280 (H-Bond), Phe279 (π-π) |
| 6c | -15.568 | Phe192, Leu259, ILeu262, Ile263, Arg265, le266, Arg269, Trp218, Phe221, Leu224, Ala225, Cys228, Phe279, Ser280, Phe295, Leu298, Ile380, Val388, Leu391, Ile403, Met407 | Arg269 and Ser280 (H-Bond), Phe279 (π-π) |
| 6d | -12.22 | Phe192, Trp218, Phe221, Leu224, Ala225, Cys228, Leu259, Leu262, Ile263, Arg265, Ile266, Arg269, Phe279, Ser280, Phe295, Leu298, Ile380, Val388, Leu391, Met399, Ile403 | Arg269 and Ser280 (H-Bond) |
| 6e | -11.569 | Phe192, Trp218, Phe221, Leu224, Ala225, Cys228, Leu259, Leu262, Ile263, Arg265, le266, Arg269, Phe279, Ser280, Gly294, Phe295, Leu298, Ile380, Val388, Leu391, Ile403, Met407 | Phe279 (π-π) |
| 6f | -8.404 | Phe192, Trp218, Phe221, Leu224, Ala225, Cys228, Leu259, Leu262, Ile263, Arg265, ile266, Arg269, Phe279, Phe295, Leu298, Val388, Leu391, Ile403 | - |
| 8a | -12.994 | Phe192, Trp218, Phe221, Leu224, Ala225, Cys228, Leu259, Leu262, Ile263, Arg265, Ile266, Arg269, Phe279, Ser280, Gly294, Phe295, Leu298, Ile380, Gly384, Val388, Leu391, Met399, Ile403 | Arg269 and Ser280 (H-Bond) |
| 8b | -14.352 | Phe192, Trp218, Phe221, Leu224, Ala225, Cys228, Leu259, Leu262, Ile263, Arg265, Ile266, Arg269, Phe279, Ser280, Phe295, Leu298, Val388, Leu391, Ile403 | Arg269 and Ser280 (H-Bond), Phe279 (π-π) |

**Repurposed the chosen ring system from histamine receptor (anti-allergic) to retinoic acid receptor *beta* (anti-cancer/inflammation) with DMPK measurement:**

Among all the synthesized compounds, based on good docking scores, some compounds were selected for RAR β agonist activity (**Figure 8**). In all the tested compounds, most positions of substituents on phenyl ring at N-1 of cyproheptadine core were well tolerated, however, substitution of –COOH group at R3 position has shown good RAR β agonist activity. Compound 6c has shown comparable agonist activity with reference compound RA (all trans-retinoic acid) and highest among all the tested compounds. Similarly, compound 8a also showed good activity as compared to rest of the tested compound. If we compare compound 6c (R1=H), 6b (R2=CH$_3$), 6d (R1=NO$_2$), and 8a (R1=NH$_2$), all have docking scores of -15.568kcal/mol, -15.324kcal/mol, -12.22kcal/mol, and -12.994kcal/mol respectively, however, a very large difference in biological activity has been observed. Moreover, it seems, the common -COOH functionality at R3 position is complementary to binding amino acid residues (i.e. Ser280, Arg269) however the presence of other substituents, and their juxtaposition in the binding pocket might have caused such large difference in biological activity.

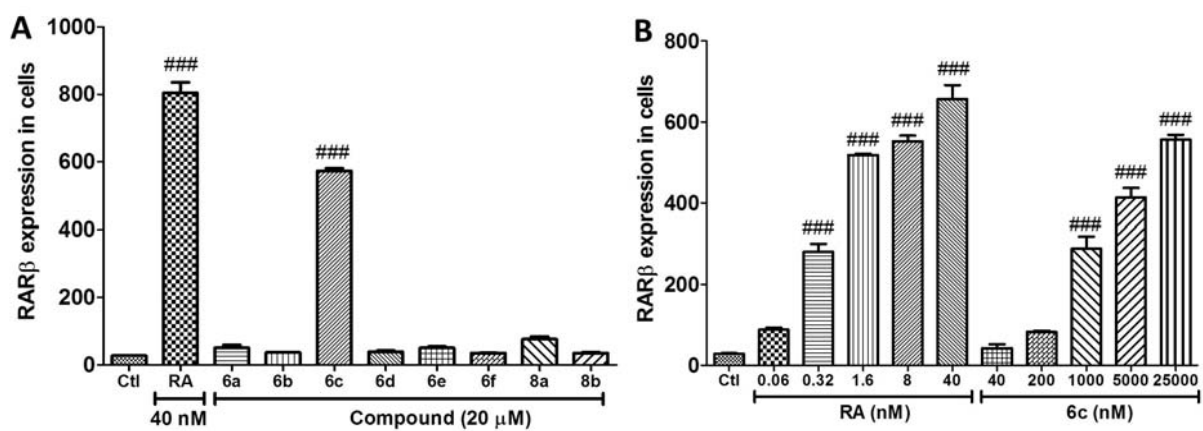

**Figure 8:** (A) The human RAR β reporter bioassay for synthesized compound, (B) The concentration dependent RAR β bioassay for RA (all trans retinoic acid) and 6c.

**Conclusion:**

Our ChOS approach made us propose the TR-dual screening to select a pair of ring system and target. The selected and identified ring system, cyproheptadine was abundantly found in ant-histamine drugs but was never considered for RAR β in literature study to present how to find a new scaffold from clinically used ring systems. This dual approach i.e. target to drug scaffold and drug scaffold to target can be combined for a novel approach in drug discovery. It is regretful that there are still majority of ring systems from FDA approved drug that are limited in use and can be exploited for drug-repurposing approach. Even though the pair of cyproheptadine and RAR β could be picked to validate our screening, the TR-dual screening can contribute to repurposing for current ring systems and future occurring ring systems (unprecedented drug scaffold).

**Experimental Section**

**1. Collection and preparation of dataset:**

The compound data containing ChEMBL Id and canonical smiles as 1,583,897 single entries was downloaded from the ChEMBL version 21 and imported into KNIME(15). The structural information carrying SMILES was converted and molecular properties were calculated by using CDK toolkit(16). All biological activity data were downloaded additionally and assigned to compound data. Similarly, the protein data containing protein number as 6,930 single entries was downloaded from the ChEMBL version 21 and imported into KNIME. From PDB bank, 219,393 ligands without hydrogen in 3D structures of deposited proteins were downloaded from the protein data bank downloader tool. As a criterion of protein filtration, protein accession id was assigned to PDB database downloaded from Uniprot database. Furthermore, to exclude protein containing marketed drug, protein data containing pharmaceutical development step information was downloaded from Therapeutic Target Database(17). The above following steps ensured in the final selection of research/clinical trial targets.

**2. Preparation of ligands:**

Input to glide docking needs compounds in 3D type format. The conversion of 2D-to-3D was accomplished with the Schrodinger LigPrep program(18). The LigPrep generates multiple states tautomers, stereoisomers, desalt, and ionization at a pH range (7 ± 2) using Epik(19). Each ligand's specified chirality were retained and generated 32 conformers unit per ligand.

All 3D conformer structures were energy minimized with the OPLS_2005 force field(20).

**3. Preparation of proteins:**

All protein structures were prepared using Protein Preparation Wizard(21). Water molecules were deleted, bond orders were assigned, hydrogens were added. Hydrogen bond was assigned using PROPKA (pH 7.0). Protein structures were refined through restrained minimization until converges within 0.30 Å RMSD for heavy atoms.

**4. Prediction of ADMET properties**

Absorption, Distribution, Metabolism, Excretion and Toxicity properties of compounds were predicted by using QikProp module of Schrodinger(22). It predicts properties such as octanol/water partition, log BB, overall CNS activity, Caco-2 and MDCK cell permeability, logKhsa for human serum albumin binding and log IC50 for HERG K+ channel blockage.

**5. Molecular Docking Simulations:**

The ligands were docked using SP (Standard Precision) and XP (Xtra Precision) protocol of Glide(23). The receptor grid had to be generated from protein structure before docking the ligands. An outer box was used for assigning ligands and an inner box was used to find a center site point of ligand in receptor grid generation. The inner boxes and outer boxes were set to the lengths of 10Å x 10Å x 10Å and original x-ray ligand respectively. The scaling factor of van der Waals radius was set to 1.0 and partial charge cutoff was set to 0.25. The centroid of the outer box was selected by considering the original x-ray ligand and ligands were docked flexibly. In docking, the default setting conditions were applied. The post-docking minimization was performed with the threshold for rejecting minimized pose and number of pose per ligand to included was set to 0.50 kcal/mol and 10 respectively.

**6. 3D similarity calculation**

3D similarity calculation was performed using OpenEye's ROCS software(24, 25). The ROCS is a popular tool because of its efficient and effectiveness. Before running ROCS, ligand structures in x-ray protein set was filtered for druggability and it was accomplished by assigning TTD database information to ligand entries and excluding entries label with successful target. Additionally, inorganic compounds and compounds with molecular weight beyond the range of 250 to 800 were excluded. Similarly, the virtual libraries were filtered on the basis of predicted molecular properties (HERG score < -5, Oral > 80%, Rule of five < 2, Rule of three < 1, and protonated amide compound was deleted). The ROCS was designed to perform the Shape based similarity search and it was conducted by comparing the shape of conformers of x-ray ligand as a reference set and ring structure as well as virtual libraries as query molecules(26,27). Shape similarity calculation was based on the Tanimoto combo score and during calculation a Tanimoto cut off score was set –0.1 to also include low score result. The maximum result number was set to 500 based on the high order of the Tanimoto combo score. For each similarity scores calculated for each x-ray ligands, the basic statistical values, such as the maximum, minimum and mean values were calculated. From shape similarity score matrix Ref Tversky Combo Max, Tanimoto Combo Mean, Ref Tversky Combo Mean, Tanimoto Combo Max, Ref Tversky Combo Min, Color Tanimoto Min, count scores were multiplied to each other, resulting in new scores (fused score) and sorted accordingly.

$$\text{Fused score} = \text{RefTverskyCombo\_Max} \times \text{TanimotoCombo\_Mean} \times \text{RefTverskyCombo\_Mean} \times \text{TanimotoCombo\_Max} \times \text{RefTverskyCombo\_Min} \times \text{ColorTanimoto\_Min} \times \text{count}$$

### 7. Target (T) & ring system (R) dual screening

In the adjusted scale of both targets and ring systems, a bi-directional TR dual screening (**r**ing system based target screening, **t**arget system based **r**ing screening) were conducted. As a preliminary screening, bi-directional docking was neither facile nor productive due to the size of the system (M: the number of PDBs on 97 targets = 3424 PDBs, N: the number of conformation of 71 ring systems = one conformer). Therefore, 3D similarity calculation of M x N matrix was expected for high speedy performance. At that time, chemo-centric assumption can permit the converted information from PDB structures of 97 targets into ligand conformers of the PDBs. The initial 3D similarity based screen helps to reduce the large pool of PDBs and feasible for bi-directional docking.

### 8. Virtual library design

The virtual library was generated using combiglide (28). A reagent database (virtual library) limited to aryl or vinyl bromide and commercial available library in ZINC database was generated from Alfa-aesar database with multiple states tautomers, stereoisomers, ionization at a pH range (7 ± 2). The generated virtual libraries were untangled and minimized.

### 9. Synthetic chemistry

**General procedure of chemistry:** All reactions were performed under nitrogen atmosphere

(anhydrous conditions). All of the reagents were purchased from Sigma Aldrich, TCI, Alfa Aeser with a purity of 95% or higher and used without further purification. All of the experimental instruments used in the reaction were thoroughly washed and then dried in a dry oven at 70° C. Anhydrous tetrahydrofuran (THF), toluene, diethyl ether (Et2O), tert-butyl methyl ether (TBME) and dimethylformamide (DMF) was purchased from Sigma Aldrich. Iridium catalysts and organolithium catalysts were stored in a refrigerator at -10° C under dark conditions. The reaction pattern was confirmed by UV irradiation, CAM stain (Cerium Ammonium Molybdate stain), PMA stain (Phospho molybdic acid stain) and Ninhydrin Stain on Silica TLC plate (60 F-254). BRUKER Ascend NMR instruments used for 1H-NMR (600 MHz) and 13C-NMR (150 MHz) spectra in CDCl3 solvent with regard to tetramethylsilane as the reference for chemical shift values. The chemical shifts (δ) values were expressed in parts per million (ppm) and coupling constants (J) in hertz (Hz).

**Synthesis of cyproheptadine derivatives and reagent library:** The selected compounds (6a-f and 8a-b) (**Table 1**) were synthesized starting from 4-(5H-dibenzo[a,d][7]annulen-5-ylidene)piperidine. The 4-(5H-dibenzo[a,d][7]annulen-5-ylidene)piperidine was obtained from commercially available cyproheptadine hydrochloride sesquihydrate (**1**), which was transformed into cyproheptadine (**2**) using 10% NaOH in ethanol. The demethylation was afforded by converting **2** into ethyl carboxylate derivative (**3**) followed by hydrolysis using KOH in ethanol to obtain 4-(5H-dibenzo[a,d][7]annulen-5-ylidene)piperidine (**4**) (**Figure 9**).

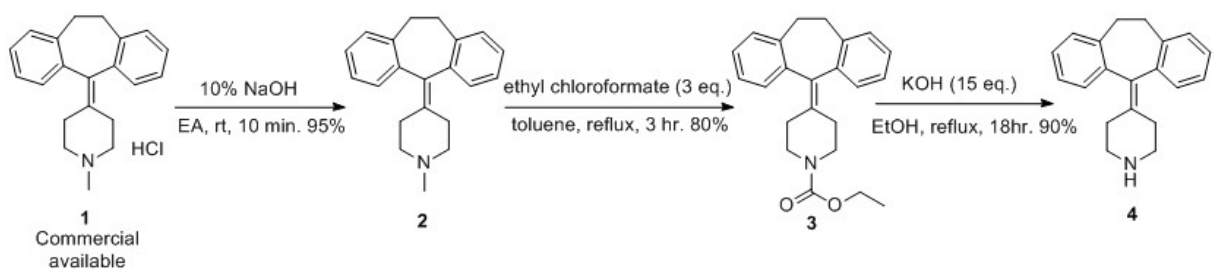

**Figure 9:** General scheme for synthesis of 4-(5H-dibenzo[a,d][7]annulen-5-ylidene)piperidine (4).

The obtained **4** is further N-arylated using various functional reagent library of methyl bromo benzoate derivative to obtain the target compound. **Figure 10** outlined the reactions and condition for synthesis of target compound and detailed procedure is provided in

**supplementary file 3**. Moreover, due to unavailability of some functional reagent library, it was prepared by performing the nitration of bromo benzoic acid (**see supplementary file 3**).

**Synthesis of reagent library and target compound:** As outlined in **Scheme S1** (**supplementary file 3**), the reagent library was prepared starting from protecting the carboxylic group through methylation using $H_2SO_4$ and MeOH as reagent and solvent. Regardless of the ortho, meta, and para positions of the carboxyl group, the reaction proceeded at a high yield. The $HNO_3$ and $H_2SO_4$ were used for nitrification. The **Figure 9** outlined the synthesis of target compound. At first step, both Sphos and palladium catalysts were used 10 times more than the conventional reaction to maintain the anhydrous condition. The obtained substituted Nitro derivatives were reduced to amine using iron, and the reaction proceeded at a high yield regardless of the reagent. Finally, the final product was completed through demethylation using KOH.

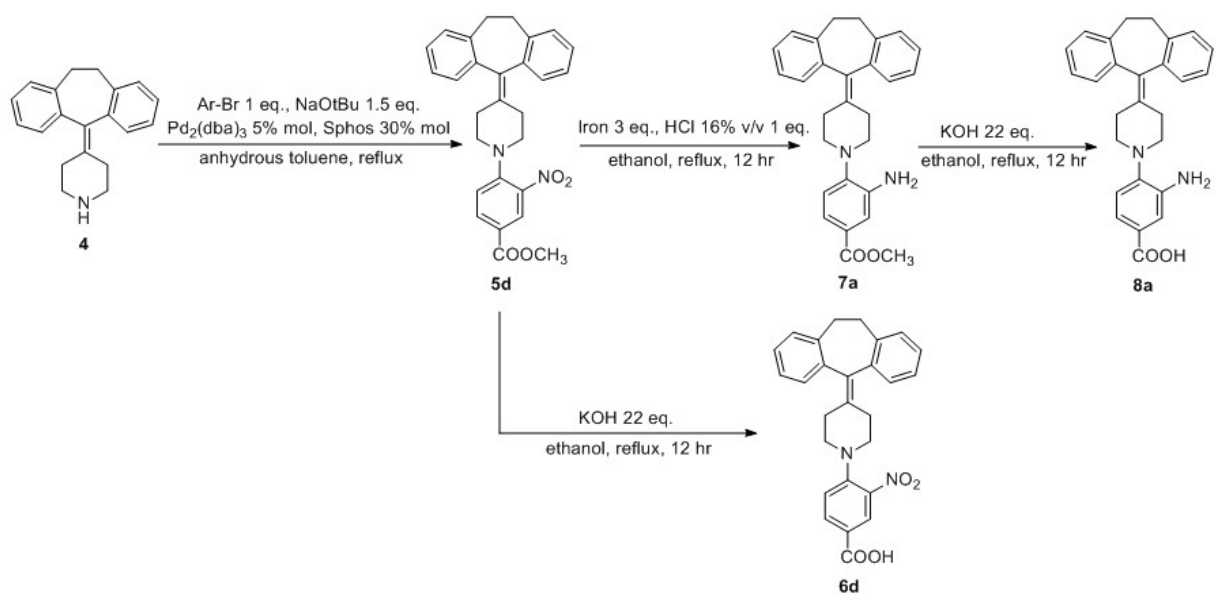

**Figure 10:** General scheme for synthesis of target compounds (6a-f and 8a-b).

## 10. Human Retenoic Acid Receptor Beta (RAR β) Reported Assay

To measure the human RAR β agonist activity for our compounds, we have performed human RAR β assay (Cat No: #IB02101-32, PA, 16801, USA) in a 3*32 assays in a 96 well plate format. The assay was performed according to the kit protocol. Cells were seeded in a 96 well plate that has been provided with kit and treated with various concentrations of negative and positive control. Treated cell plate was incubated into a 37°C, humidified with 5% $CO_2$ incubator for about 22-24 hours. On next day the supernatant was removed and the Luciferase detection reagent was added. The light emission from each assay well has been quantified using luminescence system. The RAR β bioassay result for the synthesized compound is shown in **Figure 8**.

**Conflict of interest:**

The author(s) confirm that this article content has no conflicts of interest.

**Acknowledgment:**

This study was supported by the Basic Science Research Program of the National Research Foundation of Korea (NRF), which is funded by the Ministry of Education, Science and Technology (*No.*: 2017R1E1A1A01076642). Authors would like to thank OpenEye Scientific Software providing the academic free license.

**Author contributions:**

M.-h. K. and C. J. conceived and designed the study. Under M.-h. K.'s plan, C. J. and S. K. carried out all modeling & data work and C. J. synthesized every tested compound. Under S.Y.K.' guidance, L. S. performed in vitro experiment. M.-h. K., C. J. and S.K. analyzed the data and wrote the manuscript. M.-h. K. and S.Y.K. provided the molecular modeling lab and in vitro research work facility. All authors read and approved the final manuscript.